\documentclass[12pt,english]{article}
\usepackage[T1]{fontenc}
\usepackage[latin9]{inputenc}
\usepackage{graphicx}

\newenvironment{lyxcode}
{\par\begin{list}{}{
\setlength{\rightmargin}{\leftmargin}
\setlength{\listparindent}{0pt}
\raggedright
\setlength{\itemsep}{0pt}
\setlength{\parsep}{0pt}
\normalfont\ttfamily}%
 \item[]}
{\end{list}}

\usepackage{a4}

\usepackage{babel}

\begin{document}

\title{The Concept of \\
an Emergent Cosmographic Vacuum\vspace{-1cm}}
\maketitle
\begin{lyxcode}
\begin{center}
\textrm{\textsc{Fritz~W.~Bopp}}\\
\textrm{\textsc{University~of~Siegen}}
\par\end{center}\end{lyxcode}
\begin{abstract}
The argument for an \emph{{}``Emergent} \emph{Cosmographic Vacuum''}
state which generates fermion and weak boson masses is outlined. Its
limitations and its consequences are discussed. Predictions for LHC
are presented.
\end{abstract}
The \emph{hierarchy problem} in particle physics is used as a guidance
for concepts beyond standard model physics~\cite{Ellis79}. The argumentation
often presented wrongly presumes a separation of particle physics
and cosmology. Without such a separation there is no need to directly
connect masses to GUT scale physics as a manageable scale is available
from the cosmological constant.

Of course this just changes the context as cosmology actually contains
a worse hier\-archy problem~ in its vacuum structure~\cite{weinberg89}:
The cosmological constant is taken to correspond to the vacuum energy
density caused by a condensate~\cite{Bousso07,Bousso}. The properties
of the condensate have somehow to reflect a Grand Unification scale
of the interactions when it was formed (i.e. above $10^{15}$~GeV),
whereas the flatness of the universe requires a non-vanishing but
tiny cosmological constant (about $3$~meV). 

The size of the gap rises a serious question. It is possible that
a solution of a type envisioned for the particle physics hierarchy
problem \cite{Susskind,Giudice:2007qj} will eventually be available.

Here we adhere to an opposite opinion and consider it impossible to
connect such scales. A Lagrangian with Grand Unification scale mass
terms can then not contain minima in its effective potential involving
such tiny scales. Of course, condensates contain compensating energy
terms, but without a new scale true field theoretical minima have
to stay on a GUT scale or they have to vanish%
\footnote{Studying the cosmological evolution similar ideas involving a time-dependent
vacuum state~\cite{Cai:2007us,Neupane,Krauss:2007rx} were considered.
However, based on the scale argument we here stick to the hypothesis
that the vanishing of the energy density of the true minimal state
does not dependent on the time. An undisturbed vanishing vacuum energy
is also postulated by~\cite{Volovik:2007fi,Klinkhamer:2009gm}. A
somewhat similar analogy to semiconductors is used in their argumentation.%
}. 

The observed non-vanishing cosmological constant then means that a
true minimal vacuum state is not reached. The spontaneous symmetry
breaking has to be replaced by an evolving process which is not finished.
The present condensate has to be quite close to the final minimum.
 It has to be constant over cosmic distances and it has to sufficiently
decouple from the visible world. 

What could be the history of such a physical vacuum state? In a chaotic
initial phase of GUT scale temperature tidily bound composite states
are formed with considerable statistical fluctuations. As these states
are 'massless' on a GUT scale they or weakly bound configurations
of them can reduce their remaining energy by geometrically extending.
In this way they also more and more decouple from the hotter rest.
Eventually structures are formed in a quantum mechanical process which
are constant and coherent on a sizable cosmic scale.

The advantage of this picture is that it requires no new scale. States
without scale are called {}``gap-less''~\cite{Volovik:2007fi}.
Without such a scale the evolution of the dark energy in a comoving
cell $\epsilon_{\mathrm{vac.}}$ has to be something like:\[
\partial\epsilon_{\mathrm{vac.}}/\partial(\epsilon_{\mathrm{vac.}}t)=-\kappa\epsilon_{\mathrm{vac.}}\]
where $\kappa$ is a dimensionless decay constant and $t$ the time.
It leads to a simple linear decrease. The absence of the usual exponential
decrease has the consequence that the age of the universe is no longer
practically decoupling and irrelevant. The expansion of the universe
is not linear in time. In the above equation the time $t$ has to
be replaced by the expansion parameter $a$ , i.e. $\epsilon_{vac.}\propto\frac{1}{a}$.
The expansion constant is thought to be $a\sim\sqrt{t}$ initially
and $a\sim t^{2/3}$ later on~\cite{Frieman:2008sn}. The problematic
ratio\[
{\frac{\epsilon_{GUT}}{\epsilon_{vacuum}(t_{0})}=10^{27}}\]
can then be obtained from the age of universe \[
a\propto t{_{0}}^{0.5}=(5\cdot10^{46}\frac{1}{M_{GUT}})^{1/2}\ \mathrm{resp.}\ (5\cdot10^{46}\frac{1}{M_{GUT}})^{2/3}\]
to the accuracy of the consideration. The ratio of grand unification
and present vacuum scale just connects to the age $t_{0}$ as an outside
quantity. 

It allows to understand the cosmological hierarchy problem on a conceptual
level. Of course the argumentation is rather vague and there is not
even a cosmological toy model. However, this is an intrinsic problem
of {}``Emergent'' phenomena. Without the constraint that the physical
vacuum is at an actual minimum there is too much freedom and it is
hopeless to find a realistic ab initio description. Actually this
is quite typical for most condensed matter in solid state physics
where the term Emergent Phenomena was coined for such objects~\cite{Laughlin,Liu2005}~%
\footnote{To stress the new situation the term {}``Cosmographic'' might be
useful~\cite{bopp:cosmo2,bopp:cosmo}. As in \emph{geography} many
properties of the \emph{Cosmographic Vacuum} are dependent on a chaotic
history and seemingly accidental. The name of Cosmography was first
used by Weinberg~\cite{weinberg}.%
} .

Of course this choice is extremely ugly from a model building point
of view. It actually\emph{ }leads to a murky situation: \emph{The
vacuum is largely unpredictable but predictions are necessary for
the way science proceeds}.

In this hopeless situation a tergiversate observation offers to a
certain degree an escape. The standard model contains many aspects
with broken symmetries. It is appropriate to assume that many of them
just reflect asymmetries in the accidental vacuum and that the true
fundamental physics actually is symmetric. If one accepts this esthetic
point it is actually possible to come to a number of predictions.
The basic ignorance of the vacuum keeps them on qualitative levels.

One important outcome of this argument is that the present physical
\emph{vacuum is not unique}. This has an immediate consequence. Gravity
surely can compress the non unique vacuum condensate. The distinction
between compressed dark energy and dark matter is blurred\emph{.}
A suitable compressibility can eliminate the need of dark matter altogether
and lead to \emph{an effective MoND descriptio}n~\cite{Milgrom,Bekenstein:2010pt,Knebe:2009du}.
This natural, effective theory does not require to touch fundamental
laws and it obviously has no problem with relativistic invariance.
The offset between the baryonic and dark matter component seen after
galaxy collisions~\cite{Bradac} constitutes no problem as it takes
cosmic times to rearrange the dark energy. 

Another important outcome of the argument is that the not unique vacuum
can act as a \emph{reservoir}. It is unsatisfactory to attribute the
matter-antimatter asymmetry to the initial condition of the universe.
It is widely agreed~\cite{Quinn} that no suitable, sufficiently
strong asymmetry generating process could be identified. The Cosmographic
Vacuum offers a simple way to abolish the asymmetry as the vacuum
can just contain the matching antimatter. Nothing forbids it to contain
spin- and chargeless states like Cooper pairs of two antineutrons%
\footnote{Even in the outside world roughly 1\% of the baryons are in neutron
condensates (in neutron stars).%
}. Known condensation often involves replication processes which naturally
allow to amplify tiny asymmetries over many decades. In this way initial
statistical fluctuation can be magnified to extend over cosmic regions
in the universe. A reasonable concept is that the antibaryonic condensate
is then seed to a somewhat less tidily bound mesonic cloud and to
a gluonic component known from the chiral symmetry breaking.

As these extremely extended fermionic states are practically massless
their fermi repulsion will dominate. They provide an antigravitating
contribution in the cosmological expansion~%
\footnote{Models in which the pressure in the cosmological equation is a function
of the density were considered in ~\cite{Linder:2008ya}.%
}.

The antimatter vacuum was introduced for reasons given above. It also
affects other symmetries. Whether the resulting consequences are consistent
offers a non trivial cross check. 

The vacuum state fixes the Lorentz system.\emph{ }Its tiny energy
scale leads to a huge geometrical extension and the momenta exchanged
with the vacuum have to be practically zero. Such interactions are
described with scalar, first order terms of a low energy effective
theory~\cite{Leutwyler1996et}. With such terms the Lorentz system
of the (very light) vacuum state cannot be determined. This leads
to the observed \emph{Lorentz invariance} in the outside world ~\cite{Klinkhamer:2008nr}. 

All masses have to arise from interaction with the vacuum. Their effective
couplings should be rather similar. The mass differences originate
in distinct densities. The excessive number of mass parameters is
unacceptable for fundamental physics. Here the problem is solved by
appropriately attributing them to the Emergent vacuum. The concept
then explains Hawking's postulate~\cite{Hawking}, stating that `the
various \emph{mass matrices cannot be determined from first principles}'.
The postulate doesn't preclude that certain regularities might be
identified and eventually explained~ \cite{Hall_Salem_Watari,Donoghue}. 

How do fermions obtain their masses? The relevant interaction $q_{i}+(\bar{q}_{i})_{V}\to q_{j}+(\bar{q}_{j})_{V}$
with the vacuum (\emph{V}) in the lowest perturbative order is shown
in Figure~1. Relying on the Fierz transformation we find that it
is dominated for the low momentum limit by a scalar part of the fermion-exchange
contribution between the visible world and the vacuum. %
\begin{figure}[h]
\begin{centering}
\includegraphics[scale=0.5]{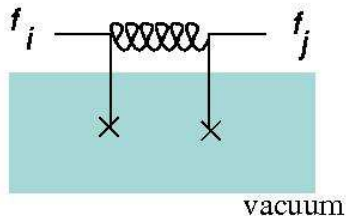}
\par\end{centering}

\caption{A process responsible for the fermion mass terms}

\end{figure}

We assume that such a flavor-dependent contribution stays important
if higher orders in the perturbative expansion are included. The matrix
elements depend on the corresponding fermion densities and on the
properties of their binding, as interactions with fermions involves
a replacement process~\cite{Anderson,Quimbay}. Multi-quark baryonic
states should be more strongly bound than the mesonic states~\cite{Froggatt:2008hc}
and fermion masses should be dominated by the less tidily bound mesonic
contribution. In this way the required dominance of the $t\bar{t}$
contribution is not disturbed by the presence of light quark antineutrons.

The\emph{ {}``flavor half-conservation''} is a serious problem in
the conventional view. Here the flavor of $q_{i}$ has not to equal
the flavor of $q_{j}$. In this way \emph{flavors conservation} can
be restored and the apparent flavor changes in the visible world can
be attributed to a reservoir effect of the vacuum. As the vacuum has
to stay electrically neutral the mass matrix decomposes in four $3\times3$
matrices which can be diagonalized and the CKM matrix can be obtained
in the usual way. As there are no new energy scales affecting the
diagonalization one finds flavor changing neutral currents are suppressed
as in the one Higgs model on a tree level. If the coherent vacuum
state is properly included unitarity relations are not affected. 

Our vacuum state is obviously not symmetric under $CP$ and $CPT$
symmetry. This allows to restore these symmetries for fundamental
physics. Without any assumptions about discrete symmetries it is then
easy to see why $CPT$ is conserved separately in our outside world
and why $CP$ not. 

In the low momentum limit a vacuum (\emph{V}) interaction of $q_{i}+(\bar{q}_{i})_{V}\to q_{j}+(\bar{q}_{j})_{V}$
will equal $\bar{q_{j}}+(\bar{q}_{i})_{V}\to\bar{q}_{i}+(\bar{q}_{j})_{V}$.
In consequence the asymmetry of the vacuum cannot be seen and $CPT$
is separately conserved in the outside world. 

On the other hand the $(\bar{q}_{i})_{V}/(q_{i})_{V}$ asymmetry in
the vacuum will differentiate between $q_{i}+(\bar{q}_{i})_{V}\to q_{j}+(\bar{q}_{j})_{V}$
and $\bar{q}_{i}+(q_{i})_{V}\to\bar{q}_{j}+(q_{j})_{V}$ . In consequence
$CP$ appears as not conserved~%
\footnote{Consider the $(K_{0},\bar{K_{0}})$ system as an example. We assume
that such a pair was produced and that the $K_{0}$ remnant is observed
in the interference region. As postulated above there are more $\bar{d}$
anti-quarks than $d$ quarks in the vacuum. The amplitude, in which
a pair of $d$ quarks is effectively deposited in the vacuum during
a $K_{0}\to\bar{K_{0}}$ transition and then taken back later on during
two $\bar{K_{0}}$ decays, obtains a different phase as that of conjugate
case of a pair of $\bar{d}$ anti-quarks deposited for the corresponding
time. This changing phase exactly corresponds to what is experimentally
observed in $CP$ violation. %
}.%
\begin{figure}[h]
\begin{centering}
\includegraphics[scale=0.5]{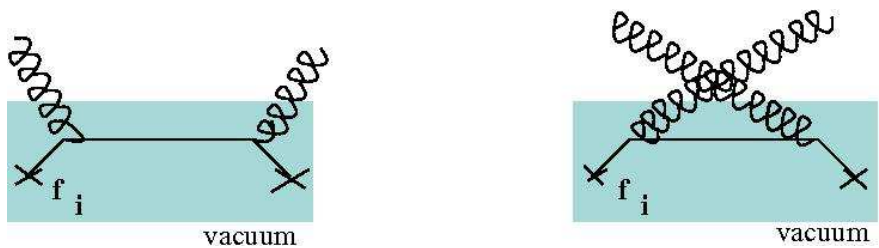}
\par\end{centering}

\caption{A process responsible for the vector boson mass terms}

\end{figure}

How do weak vector bosons obtain their masses? Relevant is a Compton
scattering process like $W_{\mu}+(\{\bar{q}\cdots\}_{i})_{V}\to W_{\nu}+(\{\bar{q}\cdots\}_{j})_{V}$
shown in Figure~2. In the low momentum limit the interaction with
gauge bosons essentially measures the squared charges of the vacuum
content.

Gluonic and Technicolor~\cite{Raby} like mesonic condensates states
are $U(1)_{B}$ neutral and cannot contribute to a $m_{B}$-mass term.
The appearance of baryonic or antibaryonic states in the vacuum provides
a $U(1)_{B}$ charge. It creates a $(B,W_{0})$ mass matrix, which
 diagonalizes in the usual way. The symmetry of this matrix and the
electrical neutrality of the vacuum ensures $m_{\gamma}=0$. The consistency
has again to be taken as a non-trivial success of the concept.

\emph{Can one make predictions for LHC}? Three {}``vacuum'' fluctuations
in bosonic densities are of course needed for the third component
of the weak vector bosons. In principle such phononic excitations
can be built with arbitrary $f_{i}\bar{f}_{j}$ -pairs and there should
be plenty of such techni-pion-like states presumably within an order
of magnitude of the Weak Boson mass. 

It is not difficult to distinguish these bosons from the usual Higgs
boson~\cite{Higgs} as they couple to the fermions in a completely
distinct way: They are private {}``Higgs'' particles~\cite{Porto},
which couple (except for the Weak Bosons) exactly to one fermion type.
There is no flavor dependence in the coupling strength of various
techni-pion like states as masses were differentiated just by distinct
densities.

If they exist $\nu\bar{\nu}$ {}``Higgs''-bosons might have the
lowest mass. Such bosons would predominantly decay into neutrinos
and its signature would be that of invisible Higgs bosons~\cite{ZhuS.-h}
with three distinct mass values. 

The light-fermion {}``Higgs''-bosons are not suppressed by tiny
coupling constants. If the energy is available they will be easy to
observe. The large-transverse-momentum-jet production at Fermilab
limits the mass of the light-fermion {}``Higgs''-bosons to an energy
above $1{\rm \,{TeV}\,}$\cite{Fermilab} and LHC will extend this
limit. 

The Emergent-Cosmographical-Vacuum concept is not a beautiful scenario.
If correct the degree to which theory can be developed is quite limited
and we can forget the dream about ever reaching the \emph{'Theory
of Everything}'. However things fit together in a surprising way on
a qualitative level and the assumptions of the Cosmographical-Vacuum
concept are not unpersuasive. Private Higgs particles with flavor
independent coupling constant would be an indication that an emergent
scenario would be \emph{nature's choice}.


\begin{thebibliography}{37}
\bibitem{Ellis79}J.~R.~Ellis, M.~K.~Gaillard, A.~Peterman and
C.~T.~Sachrajda, \textquotedbl{}A Hierarchy Of Gauge Hierarchies,''
Nucl.~Phys.~B \textbf{164} (1980) 253.

\bibitem{weinberg89}S.~Weinberg, {}``The cosmological constant
problem,''~Rev.~Mod.~Phys. \textbf{61} (1989) 1--23.

\bibitem{Bousso07}R.~Bousso, \textquotedbl{}TASI Lectures on the
Cosmological Constant,'' arXiv:0708.4231 {[}hep-th].

\bibitem{Bousso}R.~Bousso, \textquotedbl{}Precision cosmology and
the landscape'', arXiv:hep-th/0610211.

\bibitem{Susskind}L.~Susskind,   ``The Gauge Hierarchy Problem, Technicolor, Supersymmetry, And All That.   (Talk),''   Phys.\ Rept.\  {\bf 104} (1984) 181. 

\bibitem{Giudice:2007qj}G.~F.~Giudice,   ``Theories for the Fermi Scale,''   J.\ Phys.\ Conf.\ Ser.\  {\bf 110}, 012014 (2008)   [arXiv:0710.3294 [hep-ph]]. 

\bibitem{Cai:2007us}R.~G.~Cai,   ``A Dark Energy Model Characterized by the Age of the Universe,''   Phys.\ Lett.\  B {\bf 657} (2007) 228   [arXiv:0707.4049 [hep-th]]. 

\bibitem{Neupane} I.~P.~Neupane,   ``A Note on Agegraphic Dark Energy,''   Phys.\ Lett.\  B {\bf 673}, 111 (2009)   [arXiv:0708.2910 [hep-th]]. 

\bibitem{Krauss:2007rx}L.~M.~Krauss and J.~Dent, ``The Late Time Behavior of False Vacuum Decay: Possible Implications for Cosmology and Metastable Inflating States,'' Phys.\ Rev.\ Lett.\ {\bf 100} (2008) 171301 [arXiv:0711.1821 [hep-ph]].

\bibitem{Volovik:2007fi} G.~E.~Volovik,   ``From semiconductors to quantum gravity: to centenary of Matvei Bronstein,''   arXiv:0705.0991 [gr-qc]. 

\bibitem{Klinkhamer:2009gm} F.~R.~Klinkhamer and G.~E.~Volovik,   ``Towards a solution of the cosmological constant problem,''   arXiv:0907.4887 [hep-th]. 

\bibitem{Klinkhamer:2008nr}F.~R.~Klinkhamer, "Lorentz Invariance, Vacuum Energy, and Cosmology,'' arXiv:0810.1684 [gr-qc].

\bibitem{Laughlin}R.~B.~Laughlin and D.~Pines, ``The theory of everything,'' Proc.\ Nat.\ Acad.\ Sci.\ {\bf 97} (2000) 28.

\bibitem{Liu2005}Jiming~Liu, XiaoLong~Jin, Kwok~Ching~Tsui, {}``Autonomy
Oriented Computing - From Problem Solving to Complex Systems Modeling''
Springer (Heidelberg 2005)

\bibitem{Frieman:2008sn}J.~Frieman, M.~Turner and D.~Huterer, ``Dark Energy and the Accelerating Universe,'' Ann.\ Rev.\ Astron.\ Astrophys.\ {\bf 46}, 385 (2008) [arXiv:0803.0982 [astro-ph]]. 

\bibitem{bopp:cosmo2}F.~W.~Bopp, ``Is a Rich vacuum Structure Responsible
for Fermion and Weak-Boson Masses,'' arXiv:0903.1257 {[}hep-ph].

\bibitem{bopp:cosmo}F.~W.~Bopp, \textquotedbl{}A model with a Cosmographic
Landscape,'' arXiv:hep-ph/0702168.

\bibitem{weinberg}S.~Weinberg, {}``Gravitation and Cosmology: Principles
and Applications of the General Theory of Relativity ,'' John Wiley
\& Sons, (1972)

\bibitem{Milgrom}M.~Milgrom, ``MOND--a pedagogical review,'' Acta Phys.\ Polon.\ B {\bf 32} (2001) 3613 [arXiv:astro-ph/0112069].

\bibitem{Bekenstein:2010pt}J.~D.~Bekenstein, ``Alternatives to dark matter: Modified gravity as an alternative to dark matter,'' arXiv:1001.3876 [astro-ph.CO]. 

\bibitem{Knebe:2009du}A.~Knebe, C.~Llinares, X.~Wu and H.~Zhao,   ``On the separation between baryonic and dark matter: evidence for phantom   dark matter?,''   arXiv:0908.3480 [astro-ph.CO].

\bibitem{Bradac}M.~Bradac {\it et al.}, \textquotedbl{}Revealing
the properties of dark matter in the merging cluster MACSJ0025.4-1222,''
arXiv:0806.2320 {[}astro-ph]. M.~Bradac,   ``Dark matter: Revealing the invisible with 2 cosmic supercolliders the   'bullet cluster' 1E0657-56 and MACSJ0025-1222,''   Nucl.\ Phys.\ Proc.\ Suppl.\  {\bf 194} (2009) 17.

\bibitem{Quinn}H.~Quinn, ``How did matter gain the upper hand over
antimatter?,'' SLAC-PUB-13526.

\bibitem{Linder:2008ya}E.~V.~Linder and R.~J.~Scherrer, "Aetherizing Lambda: Barotropic Fluids as Dark Energy,'' arXiv:0811.2797 [astro-ph].

\bibitem{Leutwyler1996et}H.~Leutwyler, "Light quark effective theory,'' arXiv:hep-ph/9609465.

\bibitem{Raby}S.~Raby, S.~Dimopoulos and L.~Susskind, ``Tumbling Gauge Theories,'' Nucl.\ Phys.\ B {\bf 169} (1980) 373.

\bibitem{Anderson}P.~W.~Anderson, {}``Plasmons, Gauge Invariance,
and Mass'', Phys. Rev.~130, 1 (1963) 439.

\bibitem{Quimbay}C.~Quimbay and J.~Morales, \textquotedbl{}Effective
model for particle mass generation,'' arXiv:hep-ph/0702145.

\bibitem{Froggatt:2008hc}C.~D.~Froggatt and H.~B.~Nielsen, {}``Remarkable
coincidence for the top Yukawa coupling and an approximately massless
bound state,'' arXiv:0811.2089 {[}hep-ph]. 

\bibitem{Hawking}S.~W.~Hawking, \textquotedbl{}Cosmology from the
top down,'' arXiv:astro-ph/0305562.

\bibitem{Hall_Salem_Watari}L.~J.~Hall, M.~P.~Salem and T.~Watari,
\textquotedbl{}Statistical Understanding of Quark and Lepton Masses
in Gaussian Landscapes,'' arXiv:0707.3446 {[}hep-ph].

\bibitem{Donoghue}J.~F.~Donoghue, K.~Dutta and A.~Ross, \textquotedbl{}Quark
and lepton masses and mixing in the landscape'', Phys.~Rev.~D \textbf{73}
113002 (2006).

\bibitem{Higgs}P.~W.~Higgs, \textquotedbl{}Broken symmetries, massless
particles and gauge fields,'' Phys.\ Lett.\ {\bf 12} (1964). 132. 

\bibitem{Porto}R.~A.~Porto and A.~Zee, \textquotedbl{}Neutrino
Mixing and the Private Higgs,'' arXiv:0807.0612 {[}hep-ph].

\bibitem{Volovik}G.~E. ~Volovik, {}``From semiconductors to quantum
gravity: to centenary of Matvei Bronstein'', arXiv:0705.0991v3 {[}gr-qc].

\bibitem{ZhuS.-h}  S.~h.~Zhu,   `Detecting an invisibly Higgs boson at Fermilab Tevatron and CERN LHC,''   Eur.\ Phys.\ J.\  C {\bf 47} (2006) 833   [arXiv:hep-ph/0512055]. 

\bibitem{Fermilab}L. Sawyer (for the CDF and D0 Collaborations),
\textquotedbl{}Jet physics at the Tevatron'', Acta Phys. Polon. B
\textbf{36}, 417 (2005).
\end{thebibliography}
\end{document}